\magnification=1200
\def\gtsim {>\kern-1.2em\lower1.1ex\hbox{$\sim$}}
\def\ltsim {<\kern-1.2em\lower1.1ex\hbox{$\sim$}}
\def\ref {\hangindent=3em \hangafter=1 \noindent}
\def\lap{\hbox{{\lower -2.5pt\hbox{$<$}}\hskip -8pt\raise
-2.5pt\hbox{$\sim$}}}
\def\gap{\hbox{{\lower -2.5pt\hbox{$>$}}\hskip -8pt\raise
-2.5pt\hbox{$\sim$}}}
\baselineskip=16pt 
\centerline{\bf Constraints on the Strength of Primordial Magnetic Fields}
\centerline{\bf from Big Bang Nucleosynthesis Revisited}
\vskip .10in
\centerline{Baolian Cheng,$^{1}$ Angela V. Olinto,$^{1,2}$ David N. 
Schramm,$^{2}$ and James W. Truran,$^2$}
\vskip .10in
\noindent {\it $^1$ Los Alamos National Laboratory, NIS-2 MS D436, Los
Alamos, NM 87545.}
 
\noindent {\it $^2$ Department of Astronomy and Astrophysics and Enrico 
Fermi Institute, The University of Chicago, 5640 South Ellis Avenue,
Chicago, IL 60637.} 

\vskip 0.25in
\centerline {\bf ABSTRACT}
In this paper, we revisit in detail  the effects of primordial 
magnetic fields on big bang nucleosynthesis (BBN) including a 
discussion of the magnetic field geometry and the
anomalous magnetic  moment. The presence of magnetic fields affects BBN
by (1)  increasing the weak  reaction rates; (2)  increasing the
electron density due to changes to  the electron phase space; and (3)
by increasing the 
 expansion rate of the universe, due both to the magnetic  field energy
density and to the modified electron energy density. Of the effects
considered,  the  increase
in the expansion rate due to the magnetic field energy  is the most
significant for the interests of BBN.  
The   allowed magnetic field intensity  at the end of 
nucleosynthesis (0.01 MeV) is about $2 \times 10^{9}$G and corresponds 
to an upper limit on the magnetic field energy density of about 
28\% of the neutrino energy density ($\rho_B
\le 0.28 \rho_\nu$).
\vskip .25in 
\noindent PACS number(s): 98.80.Cq, 98.62.En 

\vskip .25in
\centerline {\bf 1. Introduction}

Big bang nucleosynthesis (BBN) provides an unique  quantitative window for
processes occurring in the  early universe$^{[1]}$  between  temperatures
of $ 1 {\rm MeV}$ and $0.01 {\rm MeV}$. The agreement between the light
element abundances predicted by BBN and observations strongly constrains
the dynamics of the universe at this epoch including the presence  of
strong magnetic fields.  A primeval magnetic field existing during
nucleosynthesis would have three major effects on BBN: (i) it would alter
the  weak interaction rates, (ii) it would modify the electron densities
in phase space,  and (iii) it would increase the cosmological  expansion
rate. Some  of these effects  were examined by a number of authors$^{[2]}$
and most recently by   Cheng, Schramm, and Truran$^{[3]}$, Grasso and
Rubinstein$^{[4]}$,  and Kernan et al$^{[5]}$. In this paper, we revisit
our earlier analysis$^{[3]}$  and find reasonable agreement between
subsequent work by different authors$^{[4,5]}$ and our present results.
Although some slight differences remain, the basic conclusions seem
unambiguous. Here we also show that the effects of the spatial
distribution of the magnetic fields and the  anomalous magnetic moment do
not affect significantly the results. 

\medskip
\centerline {\bf 2. Three major effects of B fields on BBN} 

Jedamzik et al$^{[6]}$ have shown that neutrino decoupling effectively
damps all magnetohydrodynamic modes up to the scales  around a tenth of 
the Hubble radius at neutrino decoupling. If $l_d$ is the largest scale 
over which the magnetic field becomes spatially homogeneous due to 
neutrino damping, then $l_d \approx 0.1 H^{-1}$ at $T\simeq1$ MeV. This 
implies that if there are magnetic fields present during BBN their 
spatial distribution is very smooth on scales  smaller than
$l_d$ and the field can be taken as constant within these scales. 

The magnetic field spatial distribution needs to be taken into account if
$l_d$ is smaller than the length scales over which reactions and mixing
occur during BBN. The relevant scale to be compared to $l_d$ here is
the largest mixing length which corresponds to the neutron diffusion 
length, $d_n$. Jedamzik and Fuller$^{[7]}$ showed that 
$d_n (1$ MeV$)\lap 1$ m while the horizon $ H^{-1}(1$ MeV) 
$\simeq 10^8$ m.   Since $l_d \gg d_n$, the magnetic field is constant 
within correlated volumes and will be taken
as constant below.\footnote{$^1$} {Recently, Grasso and Rubinstein$^{[8]}$
assumed that the magnetic field during BBN had fluctuations on the scale
of the horizon at the electroweak transition which is of the order of the
diffusion length at the end of BBN. If damping due to neutrino decoupling
were not as effective as found in Ref.[7], the spatial variations of the
magnetic field could affect  BBN outcome.}
 We also assume that the field is randomly oriented within each volume of
radius $l_B$ such that the expansion rate is not anisotropic and  that  
Robertson-Walker metric is valid. 

In an uniform magnetic field with magnitude $B$ chosen to lie along a 
$z$-axis, the dispersion relation  for an electron  propagating through 
the field is  $$E = [p_z^2 + m_e^2 + 2eBn_s]^{1\over 2} + m_e \kappa
,\eqno(2.1)$$  where $n_s= n+{1\over 2}-s_z , \ (n_s=0, 1, \, ...$), $n$
is the  principal quantum number of the Landau level and $s_z =\pm 1/2$
are  spins. $e$ is  the electron charge, $p_z$ the electron momentum,  
 $m_e$ the rest mass of the electron, and   $\kappa$ is the anomalous
magnetic moment term$^{[9]}$  for an  electron in the ground state ($n=0,
\, s_z=1/2$).  For relatively weak fields (i. e.,  $B \lap 7.575\times
10^{16}$ G),  $\kappa = -{\alpha_e\over 4\pi}{eB\over m_e^2},$  while for
stronger fields,    $\kappa =   {\alpha_e \over 2\pi} (\ln {2eB\over
m_e^2})^2,$ where $\alpha_e={1\over 137}$.  The number density of states
in the interval $dp_z$ for any given value of $n_s$ in the presence  of
magnetic field is described by$^{[10]}$   $$(2-\delta_{n_s 0}){eB\over
(2\pi)^2} dp_z.\eqno(2.2)$$  We now discuss how Eqs.(2.1) and (2.2) affect
BBN in detail. 

\smallskip
\noindent 
{\it 2.1 Weak reaction rates:} 

The weak interaction rates in a constant magnetic field 
without QED correction have been derived 
by Cheng, Schramm, and Truran$^{[11]}$.
In an expanding universe,  the B field 
evolves as $R^{-2}$,   
where $R$ is 
the scale factor of the universe. During BBN, $R \propto T_\nu^{-1}$,  
where $T_\nu$ 
is the neutrino temperature.
Let $B_i$ represent the magnetic  field at an initial temperature $T_i=1$
MeV,  
$\gamma_i=B_i/B_c$ and $\gamma=B/B_c$, where  $B_c = m_e^2/e =4.4\times
10^{13}$G is the critical field at which  quantized cyclotron states begin 
to exist. The magnetic field at any temperature  $T$ can then be  written 
as $$B=B_c \gamma_i ({T_\nu\over T_i})^2\quad {\rm or}\quad \gamma=\gamma_i 
({T_\nu\over T_i})^2.\eqno(2.3)$$   With this notation,  the rate for the
reaction $n + e^+ \to p + \bar\nu_e$, is given by $$\eqalign{\lambda_a =&
{g_V^2 (1 + 3\alpha^2) m_e^5 \gamma_i T_\nu^2 \over 4  \pi^3
T_i^2}\sum_{n_s=0}^{\infty} [2-\delta_{n_s0}] \int_{\sqrt {1+2\gamma
n_s}+\kappa}^\infty {d\epsilon (\epsilon -\kappa) \over
\sqrt{(\epsilon-\kappa)^2-(1+2\gamma n_s)}}\cr
 &\qquad \qquad \times {1\over (1+e^{\epsilon Z_e + \phi_e})}  {(\epsilon +
q)^2 e^{(\epsilon + q) Z_\nu + \phi_\nu} \over (1+e^{(\epsilon + q) Z_\nu +
\phi_\nu})},\cr}\eqno(2.4)$$ where $g_V^2 (1 + 3 \alpha^2) m_e^5 /2 \pi^3 
\simeq 6.515 \times 10^{-4} {\rm sec}^{-1},$  
$g_V = 1.4146\times 10^{-49}$erg 
${\rm cm}^3$, and $\alpha = {g_A / g_V} \simeq -1.262.^{[12]}$  

Similarly, for the reaction $n + \nu \to p +
e^-$, the rate is $$\eqalign{\lambda_b =& {g_V^2 (1 + 3 \alpha^2) m_e^5
\gamma_i T_\nu^2\over 4 \pi^3T_i^2 } \sum_{n_s=0}^{\infty} [2-\delta_{n0}]
\int_{\sqrt {1+2\gamma n_s}+\kappa}^\infty {d\epsilon (\epsilon-\kappa) \over
\sqrt{(\epsilon-\kappa)^2-(1+2\gamma n_s)}}\cr &\qquad \qquad \times {1\over
(1+e^{\epsilon Z_e + \phi_e})} {(\epsilon - q)^2 e^{\epsilon Z_e + \phi_e}
\over (1+e^{(\epsilon - q) Z_\nu - \phi_\nu})}\cr &- {g_V^2 (1 + 3 \alpha^2)
m_e^5 \gamma_i T_\nu^2 \over 4 \pi^3 T_i^2} \sum_{n_s=0}^{n_{s{\rm max}}}
[2-\delta_{n_s0}] \int_{\sqrt{1+2\gamma n_s}+\kappa}^q {d\epsilon
(\epsilon-\kappa)\over\sqrt{(\epsilon-\kappa)^2-(1+2\gamma n_s)}}\cr &\qquad
\qquad \times {1\over (1+e^{\epsilon Z_e + \phi_e})} {(\epsilon - q)^2
e^{\epsilon Z_e + \phi_e} \over (1+e^{(\epsilon - q) Z_\nu - \phi_\nu})},\cr}
\eqno(2.5)$$  and for the reaction $n \to p + e^- + \bar\nu_e$, we have
$$\eqalign{\lambda_c=& {g_V^2 (1 + 3 \alpha^2) m_e^5 \gamma_i T_\nu^2 \over 4
\pi^3 T_i^2} \sum_{n_s=0}^{n_{s{\rm max}}} [2-\delta_{n_s0}]
\int_{\sqrt{1+2\gamma n_s}+\kappa}^q {d\epsilon (\epsilon -\kappa)\over
\sqrt{(\epsilon -\kappa)^2-(1+2\gamma n_s)}}\cr &\qquad \qquad \times {1\over
(1+e^{\epsilon Z_e + \phi_e})} {(q-\epsilon)^2 e^{\epsilon Z_e + \phi_e }
\over (1+e^{(q-\epsilon) Z_\nu + \phi_\nu})}.\cr} \eqno(2.6)$$ The total weak
reaction  rates for the  conversion of neutrons to protons is simply  the sum
of the above rates   $$\eqalign{\lambda_{n\,\to\,p} &= {g_V^2 (1 + 3 \alpha^2)
m_e^5 \gamma_iT_\nu^2 \over 4 \pi^3 T_i^2} \sum_{n_s=0}^{\infty}
[2-\delta_{n_s0}]\times \int_{\sqrt {1+2\gamma n_s} +\kappa}^\infty d\epsilon
{(\epsilon-\kappa) \over 1+e^{\epsilon Z_e + \phi_e} } \cr             &
\times {1\over {[(\epsilon-\kappa)^2-(1+2\gamma n_s)]^{1\over
2}}}\bigr[{(\epsilon+q)^2 e^{(\epsilon+q) Z_\nu  + \phi_\nu} \over
1+e^{(q+\epsilon) Z_\nu + \phi_\nu}} + {(\epsilon-q)^2 e^{\epsilon Z_e +
\phi_e}\over 1+e^{(\epsilon-q)Z_\nu - \phi_\nu}}\bigr ].\cr} \eqno(2.7)$$  

The parameters used above are defined as  
$\epsilon = {E\over m_e },\, q={m_n - m_p\over m_e},\, Z_e={m_e \over T_e},$
$Z_\nu = {m_e \over T_\nu}, \, \phi_e = {\mu_e \over T_e},$ and $ \phi_\nu =
{\mu_\nu \over T_\nu},$  where $m_n$ and $m_p$ are the rest masses of the
neutron and proton respectively, $T_{e}$ is the temperature of the electrons,
$\mu_i (i=e,\,\nu)$ is the chemical potential of the electron or neutrino, 
and $\phi_i (i=e,\,\nu)$ is the degeneracy parameter.

For $\beta$-decay processes to occur in the presence of a magnetic field, 
the quantum number $n_{s}$ has to satisfy 
$$\sqrt{1+2\gamma n_s} +\kappa \le q, \quad {\rm or} \quad n_s\le 
n_{s{\rm max}} = {\rm Int}\bigg [{(q-\kappa)^2
 -1 \over 2\gamma}\bigg ],\eqno(2.8)$$  
where $n_{s{\rm max}}$ is the largest integer in 
${(q- \kappa)^2 -1 \over 2\gamma}$. (Note 
that our expressions for the total rate (2.7) here is 
different from that given in Ref [4].)  

The inverse total reaction rate of the conversion of protons to neutrons is 
computed as $$\lambda_{p\to n} = e^{-qZ_e} \lambda_{n\to p}. $$ 

In order to better elucidate the effects of the field on the rates, we 
calculate analytically the variations of the reaction rates with respect to
changes in the field, and we obtain  
$$\lim_{\gamma \to 0}{d\lambda_a \over d\gamma}
\propto \sum_{n_s=0}^{\infty} [2-\delta_{n_s0}] \int_0^\infty dk f_+(\epsilon)
u_+(\gamma,\epsilon) >0,\eqno(2.9) $$  
$$\lim_{\gamma \to 0}{d\lambda_b 
\over
d\gamma} \propto \sum_{n_s=0}^{\infty} [2-\delta_{n_s0}] \int_0^\infty dk
f_-(\epsilon) u_-(\gamma,\epsilon) -   \sum_{n_s =0}^{n_{\rm smax}}
[2-\delta_{n_s0}] \int_0^q dk  f_-(\epsilon) u_-(\gamma,\epsilon)
>0,\eqno(2.10) $$ 
$$\lim_{\gamma \to 0}{d\lambda_c \over d\gamma} \propto
\sum_{n_s =0}^{n_{\rm smax}} [2-\delta_{n_s0}] \int_0^q dk f_-(\epsilon)
u_-(\gamma,\epsilon)   >0,\eqno(2.11) $$  
and  $$\lim_{\gamma \to
0}{d\lambda_{n\to p} \over d\gamma}\propto \sum_{n_s=0}^{\infty}
[2-\delta_{n_s0}] \int_0^\infty dk [f_+(\epsilon)u_+(\gamma,\epsilon) +
f_-(\epsilon)u_-(\gamma,\epsilon)] >0, \eqno(2.12)$$  
where 
$$\epsilon = (k^2 + 1 + 2\gamma n_s)^{1/2} + \kappa ,\qquad k={p_z^2\over
 m_e^2}, $$ 
$$f_\pm (\epsilon) = {(\epsilon \pm q)^2 \over (1+e^{\pm(\epsilon Z_e + 
\phi_e)}) (1+e^{-[(q \pm \epsilon) Z_\nu + \phi_\nu]})}.$$ 

$$u_\pm(\gamma,\epsilon) = 1 + {2\gamma n_s \over \epsilon ({\epsilon 
\pm q})}
\mp {\gamma n_s Z_e \over \epsilon}{e^{\pm (\epsilon Z_e+ \phi_e)}\over
1+e^{\pm (\epsilon Z_e + \phi_e)}} \pm {\gamma n_s Z_\nu \over
\epsilon({1+e^{- [(q \pm \epsilon) Z_\nu + \phi_\nu] }})}.  $$   

We also
computed Eqs.(2.9 -2.12) numerically, for various of $\gamma_i$ and $T$. 
Both calculations show that,  independent of the temperature, the presence
of a magnetic field does increase all  the weak reaction rates, including
the total neutron depletion rate.   This result is consistent with the
findings in  our previous works and with Grasso and Rubinstein's
recent calculations,$^{[4]}$ but inconsistent   with Kernan's recent
statement$^{[5]}$ that  the rates of 2-2 processes decrease as  the field
increases.     At very high temperatures $T\gg 2.5$ MeV, 
such effects are insignificant because  
the inverse reaction rates also increase with the field and are not much 
suppressed by the factor   $\exp(-qZ_e)$.  
When the temperature drops  to a point where the 
reactions $n + e^+\to p + \bar\nu$ and $n + \nu \to p + e^-$ begin to
freeze out and the neutron $\beta$-decay process dominates, then 
the total rate increases  with the magnetic field. 
However, if the  primeval field is 
not strong enough to begin with, then as the universe expands, 
it  becomes
too weak to affect the reaction rates at low temperatures. 
Our numerical calculations reveal that 
for the magnetic  field to have significant  impact on the 
reaction rates,  $\gamma_i\gap  10^3 $ or $B_i\gap 4.4\times 10^{16}$ G at 
1 MeV. 
As we discuss below,  the effect due to the
change in expansion rate is already significant for $\gamma_i\gap 10$ and
dominates over the change on the reaction rates. 

\smallskip
\noindent 
{\it 2.2 Electron density phase space} 

In a magnetic field, the phase space and energy density of electrons are
modified.  The number density and energy density of electrons ($n_e$ and 
$\rho_e$)  over phase
space as a  function of magnetic field strength are given by  $$n_e = 2{m_e^3
\gamma_i T_\nu ^2\over (2\pi)^2T_i^2} \sum_{n_s=0}^{\infty} 
(2-\delta_{n_s0})\int_{\sqrt {1+2\gamma n_s}+\kappa}^\infty d\epsilon
{(\epsilon -\kappa) \over \sqrt{(\epsilon-\kappa)^2-(1+2\gamma n_s)}}{1\over
1+e^{\epsilon Z_e + \phi_e}}\eqno(2.13)$$  and   $$\rho_e(B)=2{m_e^4\gamma_i
T_\nu ^2\over (2\pi)^2T_i^2} \sum_{n_s=0}^{\infty}(2-\delta_{n_s0})\int_{\sqrt
{1+2\gamma n_s}+\kappa}^\infty d\epsilon {\epsilon (\epsilon -\kappa) \over
\sqrt{(\epsilon-\kappa)^2-(1+2\gamma n_s)}}{1\over 1+e^{\epsilon Z_e +
\phi_e}}.\eqno(2.14)$$    Correspondingly, the pressure of electrons is 
$$P_e
= 2{m_e^4 \gamma_i T_\nu ^2\over (2\pi)^2T_i^2}
\sum_{n_s=0}^{\infty}(2-\delta_{n_s0})\int_{\sqrt {1+2\gamma
n_s}+\kappa}^\infty d\epsilon {(\epsilon -\kappa)\over 3\epsilon} {\sqrt
{(\epsilon-\kappa)^2-(1+2\gamma n_s)}\over 1+e^{\epsilon Z_e +
\phi_e}}.\eqno(2.15)$$  These expressions will reduce to$^{[10]}$   $$n_e =
{m_e^3\over \pi^2}\int_1^\infty d\epsilon {\epsilon \sqrt{\epsilon^2-1} \over
1+e^{\epsilon Z_e + \phi_e}},\eqno(2.16a)$$  $$\rho_e = {m_e^4\over \pi^2}
\int_1^\infty
d\epsilon {\epsilon^2 \sqrt{\epsilon^2-1}\over 1+e^{\epsilon Z_e +
\phi_e}},\eqno (2.16b)$$  and  $$P_e = {m_e^4\over 3\pi^2}\int_1^\infty 
d\epsilon
{(\epsilon^2-1)^{3/2} \over 1+e^{\epsilon Z_e + \phi_e}}\eqno (2.16c)$$  
if the magnetic
field is absent.  

The dependences of $n_e,\, \rho_e,\, {\rm and} \,\, P_e$  on the magnetic 
field  can be seen analytically to be  $$\lim_{\gamma \to 0}{dn_e\over
d\gamma}\propto \sum_{n_s=0}^{\infty} (2-\delta_{n_s0}) \int_{0}^\infty dk {1
\over 1+e^{\epsilon Z_e + \phi_e}}[1-{\gamma n_s Z_e\over \epsilon (1 +
e^{-(\epsilon Z_e + \phi_e)})}]>0,\eqno(2.17)$$ $$\lim_{\gamma \to 0} 
{d\rho_e
\over d\gamma}\propto \sum_{n_s=0}^{\infty}  (2-\delta_{n_s0})\int_{0}^\infty
dk {\epsilon \over (1+e^{\epsilon Z_e + \phi_e})} [1+ {\gamma n_s\over
\epsilon^2 } - {\gamma n_s Z_e \over \epsilon (1 + e^{-(\epsilon Z_e +
\phi_e)}))}] >0,\eqno(2.18)$$  and  $$\lim_{\gamma \to 0}{dP_e\over
d\gamma}\propto \sum_{n_s=0}^{\infty} (2-\delta_{n_s0}) \int_{0}^\infty dk
{k^2 \over \epsilon (1+e^{\epsilon Z_e + \phi_e})} [1- {\gamma n_s\over
\epsilon^2} - {\gamma n_s Z_e \over \epsilon (1 + e^{-(\epsilon Z_e +
\phi_e)})}] >0.\eqno(2.19) $$ where we assumed non-degenerate 
neutrinos ($\phi_\nu=0$).  These expressions indicate that, in the presence 
of magnetic
fields, due to the large  Landau excitation energy and the  decreased
cross-sectional area of each Landau level,  all of the electron thermodynamic
quantities, such as   $n_e, \, \rho_e,\, {\rm and}\, P_e$,  increase with
increasing field strengths. This,  in turn, causes a  decrease in all of the
weak  interaction rates and changes the temperature-time relationship  in BBN
calculations. Furthermore, it results in an  increase in the final neutron to
proton ratio and the abundances of the light  elements. Such an effect
becomes   significant for $\gamma_i \gap 10^{3}$ and
is sub-dominant to the effect discussed below.  

\smallskip
\noindent 
{\it 2.3 Effects on the Expansion Rate} 

The expansion rate of our universe 
is given by
$$H\equiv {1\over R} {dR\over dt} = \sqrt{ {8 \pi G \over 3}\rho }, 
\eqno(2.20)$$
where $G$ is the gravitational constant and 
$\rho$ is the total energy density. This can be  
 expressed as  
$\rho = \rho_\gamma + \rho_e + \rho_\nu + \rho_b + \rho_B,$  
where 
$ \rho_e = \rho_{e^-} + \rho_{e^+},$ 
$\rho_\nu = \rho_{\nu_{e}} +  \rho_{\nu_\mu} + \rho_{\nu_\tau}
+\rho_{\bar\nu_e} + \rho_{\bar\nu_\mu} + \rho_{\bar\nu_\tau},$ 
and the subscripts $\gamma, \,\, e,\,\, \nu_e,\,\,\nu_\mu,\,\,\nu_\tau,\,\, 
b, \,\, {\rm and}\,\, B$ stand for 
photons, electrons, $e$-neutrinos, $\mu$-neutrinos, $\tau$-neutrinos, 
baryons, and magnetic fields.

The presence of magnetic fields alters the expansion rate by the added 
energy  density of  the magnetic field 
$$\rho_B={B_c^2\over 8\pi}\gamma_i^2 \left( {T_{\nu} \over T_i} \right)^4 
\eqno(2.21)$$
and the change in the electron energy density which we can write as   
$$\rho_e \equiv \rho_e(B=0)
 + \delta \rho_e \eqno(2.22)$$

During nucleosynthesis, $e^+e^-$ annihilation transfers entropy to the 
photons but not
to the decoupled neutrinos. The neutrino temperature then follows $T_{\nu} 
\propto
R^{-1}$, i.e.,
$$ {dT_{\nu}\over dt} = - H T_{\nu} \eqno(2.23) $$
while the photon temperature satisfies
$${dT_{\gamma}\over dt} = - 3 H {\rho_e + \rho_\gamma + P_e + P_\gamma\over
d\rho_e/dT_{\gamma} + d\rho_\gamma /dT_{\gamma}}\  .\eqno(2.24) $$ 
These equations are solved simultaneously since
$H$ is a function of both $T_{\nu}$ and $T{\gamma}$, i.e., 
$$\rho(T_{\gamma}, T_{\nu}) = \rho_{\nu}(T_{\nu}) + \rho_e(T_{\gamma}, 
B(T_{\nu})) +
\rho_{\gamma}(T_{\gamma}) +  \rho_{B}(T_{\nu}) + \rho_b(T_{\gamma}) \ . 
\eqno(2.25) $$

If we define 
$$\rho_0\equiv \rho(B=0) ,\,\,\theta 
\equiv {\delta\rho_e \over \rho_0},\,\, \chi \equiv {\rho_B \over  \rho_0}
 , \eqno(2.26)$$  
we can estimate the effect of $\theta$ and $\chi$ on the time-temperature 
relation away from $e^+e^-$ annihilation:
$$T_{\gamma} \sim 10^9 K \xi \,(1 + \theta +\chi)^{- 1/4} t^{-1/2}\ . 
\eqno(2.27)$$ 
 $\xi = 4.7 $ for three types of neutrinos.  
If there is no magnetic field, Eq.(2.27) reduces to the formula in standard
BBN$^{[12]}$  $$T_{\gamma} \approx 10^9 K \xi t^{-1/2} \, .$$ 

The modified time-temperature relationship (2.27) suggests that 
the contributions of the primordial magnetic field from both the field
energy  density and the electron energy density, accelerate the expansion
rate   of the universe and decrease the time scale over which  BBN can
occur.  In particular, the neutrons will have less time to decay
 to protons than  
in the field-free case, which leads to an enhanced final $n\over p$ ratio 
and ultimately elevates the  abundance of $^4He$. 

Comparing the energy density from the electrons  
to that directly from magnetic fields, we find that for $\gamma_i <10$, 
the contribution from electrons is somewhat greater than that from the
field,   but it is still too small to be interesting with respect to 
the total energy density in the field 
free case. When the  B field is stronger, $\gamma_i > 10$, the 
contribution from the magnetic field exceeds that of the electron 
phase space and  
dominates the total energy density.  
 
\medskip
\centerline {\bf 3. Limits on the field strength from BBN}

We considered all three effects discussed above in our numerical
calculations to set a limit on the field strength allowed by BBN 
considerations. As expected, our calculations  
reveal that the abundances of the light elements are 
manifestly affected by  strong magnetic fields  ($B_i\gap 10^{15}$G).
 Although the three effects  are important for
fields in excess of $B_c$, the dominant process for setting an upper
limit on the magnetic field during BBN  is the change in the
time-temperature relation discussed in \S 2.3 in agreement with the
results of Refs.[4]  and [5].

Our numerical calculations show that for an initial magnetic field 
$B(1$ MeV) $\lap 10^{15}$G, the impact on the 
neutron to proton ratio from the 
magnetic field energy density, which decreases the neutron population,   
is more significant than the other effects.  
By using the observed abundance of $^4$He,  D, and  $^3$He, we find a
constraint on the strength of a primordial magnetic field  
which  is
equivalent to an increase in the number of neutrino families. 
To explicitly calculate these effects,  we set the
neutron lifetime $\tau_n= 887$ s  $\pm$ 2 s,$^{[13]}$ the number of neutrino
species  $N_\nu = 3 $, and compute the primordial abundances
numerically.  Similarly to the bound on $N_\nu$, the constraint on the
magnetic field energy relies on the lower limit to $\eta$  and the upper
limit to $^4$He. We use the D + $^3$He lower bound on the
baryon-to-photon ratio ($\eta \ge 2.5 \times  10^{-10}$) and an upper limit
to the $^4$He abundance ($Y_P \le 0.245$)$^{[14]}$ and find that $
\gamma_i \le 85$. This implies that the allowed magnetic field at the end of
BBN ($T_\gamma = 0.01$ MeV) is less than about $2 \times 10^{9}$G
which corresponds to a limit on the energy density of magnetic fields during
BBN  $\rho_B \le 0.28 \rho_\nu$.

\medskip
\centerline {\bf 4. Conclusion}  

In previous sections, we have provided a detailed analysis of the three 
major effects of a primordial magnetic field on 
the final abundances of the elements formed in big-bang nucleosynthesis.  
We have found  
that of the three major effects - (a)  increased  
weak interaction rates; (b)  enhanced electron densities  
in phase space; and  (c) an increased expansion rate of the universe 
by the energy densities of magnetic field and electrons - the latter effect   
dominates over  the modifications arising from 
the first two effects when 
$B_i\le 10^{15}$G even when the electron magnetic moment is included.
We have computed these effects numerically and obtained  
a revised upper limit on the allowed strength of a primordial magnetic 
field on scales smaller than $l_d$.
Our results show that, in the framework of standard big bang nucleosynthesis,
the maximum strength of a primordial magnetic field is such that 
$\rho_B
\le 0.28 \rho_\nu$.  

\vskip .20in
\centerline {\bf Acknowledgments}
B. C. would like to thank Gordon Baym for very
helpful discussions and the  Director's Fellow fund 
at Los Alamos National Laboratory for financial support. D.N.S. 
would like to acknowledge useful interactions with Hector Rubinstein. 
This research has  also been supported by grants from NASA through grant
NAGW 1321, DOE through grant  DE
FG0291 ER40606, and NSF through grants AST-93-96039 and AST-92-17969 at the
University of Chicago, and  by the DOE and by NASA through grant NAGW 2381 
at Fermilab.

\vskip .30in 

\centerline{\bf REFERENCES}
\vskip .20in
\par \ref
[1] D. N. Schramm and R. V. Wagoner, Ann. Rev. Nucl. \& Part. Sci., (1977);
T. P. Walker, G. Steigman, D. N. Schramm, K. A. Olive, and H-S Kang,
Astrophysics. J., 376 (1991) 51. P.24; 
G. Steigman, D. N. Schramm, and J. Gunn, Phys. Lett. B 66 (1977) 202; J. 
Yang, 
M. S. Turner, G. Steigman, D. N. Schramm, and K. A. Olive, Astrophys. J., 
281 
(1984) 493; 
P. J. E. Peebles, Astrophys. J., 146 (1966) 542; R. V. Wagoner, 
W. A. Fowler, and
F. Hoyle, Astrophys. J., 148 (1966) 3. 
\par \ref
[2] J. D. Barrow, Mon. Not. R. Astro. Soc., 175 (1976) 379; R. F. O'Connell 
and J. 
J. Matese, Nature, 222 (1969) 649.
\par \ref
[3] B. Cheng, D. N. Schramm, and J. W. Truran, Phys. Rev. D, 49 (1993) 5006.
\par \ref
[4] D. Grasso and H.R. Rubinstein, Astroparticle Phys. 3 (1995) 95.
\par \ref 
[5] P.J. Kernan, G.D. Starkman, and T. Vachaspati, astro-ph/9509126. 
\par \ref 
[6] K. Jedamzik, V. Katalinic, and A.V. Olinto, in preparation
\par \ref 
[7] K. Jedamzik and G. M. Fuller, Astrophys. J., 423, (1994) 33.
\par \ref 
[8] D. Grasso and H. Rubinstein, astro-ph/9602055
\par \ref 
[9] J. Schwinger, Particles, sources and fields, Vol 3, Chapter 5.6, 
Addison-Wesley, 
Redwood City, CA, 1988.
\par \ref
[10] L.D. Landau and E.M. Lifshitz, Statistical mechanics, Clarendon Press, 
Oxford, 
1938.  
\par \ref
[11] B. Cheng, D. N. Schramm, and J. W. Truran, Phys. Lett. B, 316 (1993) 
521.
\par \ref
[12] R. Wagoner, Astrophys. J. Suppl. series 162 Vol 18 (1969) 247.
\par \ref 
[13] E. Skillman, {\it et al.}, Astrophys. J., 1993, in press; also, in 
Proc. 
16th TEXAS/PASCOS Meeting, Berkeley, 1992.
\par \ref
[14] C. J. Copi, D. N. Schramm, and M. S. Turner, Phys. Rev. Lett. 75
(1995) 3981.

\vfill
\eject

\end